\documentclass[epj]{svjour}

\usepackage{graphics}
\usepackage{textcomp}
\begin{document}
\title{The secondary structure of RNA under tension}
\author{M.~M\"uller \and F.~Krzakala \and M.~M\'ezard}
\offprints{muller@ipno.in2p3.fr}
\institute{Laboratoire de Physique Th\'eorique et Mod\`eles Statistiques,
b\^at. 100, Universit\'e Paris-Sud, F--91405 Orsay, France.}
%
%

\newcommand \be {\begin{equation}} 
\newcommand \ee {\end{equation}}

\abstract{
We study the force-induced unfolding of random disordered RNA or
single-stranded DNA polymers. The system undergoes a second order phase transition from a collapsed globular phase at low forces to an extensive necklace phase with a macroscopic end-to-end distance at high forces. At low temperatures, the sequence inhomogeneities modify the critical behaviour. We provide numerical evidence for the universality of the critical exponents which, by extrapolation of the scaling laws to zero force, contain useful information on the ground state ($f=0$) properties. This provides a good method for quantitative studies of scaling exponents characterizing the collapsed globule. 
In order to get rid of the blurring effect of thermal fluctuations we restrict ourselves to the groundstate at fixed external force. We analyze the statistics of rearrangements, in particular below the critical force, and point out its implications for force-extension experiments on single molecules. 
\PACS{
      {87.14.Gg}{DNA, RNA}   \and
      {87.15.-v}{Biomolecules: structure and physical properties} \and
			{64.60.-i}{General studies of phase transitions}
     }
} 
\maketitle
\section{Introduction}
\label{intro}

The last years have whitnessed a lot of progress in the experimental study of force-induced unfolding of biomolecules using techniques such as atomic force microscopy and optical tweezers. A variety of experiments on RNA and single- or double-stranded DNA have been devised \cite{BustamanteSmith96,Liphardt02,Rief97,Rief99,BockelmannEssavez97,BockelmannThomen02,Maier00} to study the behaviour of these polymers under an external force which allows to determine their elastic properties and provides new insight into the folding problem of biomolecules. In particular, these techniques offer a means to investigate the energy landscape and folding pathways and to extract specific thermodynamic parameters. In this paper, we will study the force-induced unfolding of RNA or single-stranded DNA, with special emphasis on the effects of sequence heterogeneity (disorder).

Several theoretical models to describe molecules under external forces have been investigated for the case of general heteropolymers \cite{GeisslerShakhnovich02a,GeisslerShakhnovich02b,LeeVilgis01,KlimovThirumalai01}, polyelectrolytes \cite{VilgisJoanny00}, DNA-unzipping \cite{LubenskyNelson00,LubenskyNelson02,Marenduzzo01,CoccoMonasson01} and unfolding of RNA \cite{ChenDill00,GerlandBundschuh01}. Within a mean-field treatment of heteropolymers \cite{GeisslerShakhnovich02a,GeisslerShakhnovich02b}, disorder has been shown to be relevant in the glassy low force regime where a random energy model applies. In a critical region around the (first order) denaturation transition the disorder induces a necklace structure in the polymer chain which is intermediate between the globular and the fully extended state. The breaking of individual globular domains upon increasing the force leads to step-like force-extension characteristics \cite{LeeVilgis01}.

In the unzipping of a DNA double strand sequence, heterogeneity has been shown to modify the critical behaviour of the opening phase transition \cite{LubenskyNelson00,LubenskyNelson02} which has been characterized in terms of the statistics of elementary unzipping events below the threshold force.

The case of unfolding of RNA is more involved in that the ground state usually has a much more complicated structure than a single hairpin (the equivalent of a DNA double strand) and competes with a large number of low-lying metastable states into which the system can be driven by the external force. On the other hand, for the same reason, the force-extension characteristics at low enough forces may reveal more information about the energy landscape of the molecule \cite{ChenDill00,KlimovThirumalai01}. In previous articles, force-extension curves have been discussed for a homogeneous model of RNA \cite{MontanariMezard01,Muller02} as well as for disordered RNA \cite{GerlandBundschuh01}. The homogeneous description exhibits a second order phase transition from a collapsed globular state to an extended necklace-like phase. We will show that the phase transition persists under the introduction of sequence disorder. However, at low temperatures, the critical behaviour is modified and the globular phase becomes glassy which manifests itself in the statistics of jumps in force-extension curves. At higher temperatures, the homogeneous description has been argued to be a valid coarse-grained description of random RNA \cite{Muller02}.

The extended phase is not very sensitive to disorder, and random RNA exhibits rather few specific features in pulling experiments \cite{Maier00,GerlandBundschuh01}. Indeed the fit of the experimental data in \cite{MontanariMezard01} based on a homogeneous model is remarkably good. The force-extension curves of naturally selected RNA are usually richer than those of random sequences of the same size which probably reflects the bias of naturally selected RNA towards sequences with a particularly stable ground state and favourable folding characteristics \cite{Higgs93,GerlandBundschuh01}. 

In this paper we concentrate on the low forces and the critical
 window around the above-mentioned second order phase transition. This regime exhibits a lot of interesting features that are intimately related to
the energy landscape of the molecule. It is in this regime that the most
complex force-induced rearrangements in the secondary structure take place.
These are expected to proceed via a slow activated dynamics giving rise to
long equilibration times and ageing or hysteresis effects (see
\cite{BockelmannThomen02} for related experiments done on DNA). The latter are
probably relevant for the behaviour of large RNA molecules such as messenger
RNA or heteronuclear RNA. Most current experiments however usually operate at
forces comparable to the threshold force to unzip double-stranded DNA ($\approx 10 pN$) which is
considerably larger than the critical force mentioned above. In sufficiently
small molecules, these experiments allow for the identification of parts of
the folding pathway in short molecules \cite{Rief97,ChenDill00,Liphardt02}.
The nature of the events that dominate the force-extension curve in the
extensive phase is however rather simple in that they are transistions between
two competing foldings of a relatively small domain. The results of this
article show that the large scale energy landscape should be studied at forces
in the critical window and at modest chain extensions, and we give qualitative
predictions for the expected behaviour of random disordered molecules in this
regime.

The relative smoothness of the force-extension curves of RNA hides a large
part of the structural transitions in the molecule and thus hinders the
understanding of the underlying processes. The authors of
\cite{GerlandBundschuh01} trace the origin of the smoothness to essentially three factors: {\it
i)} the thermal effects in the form of entropic elasticity that naturally
limit the resolution of any force-extension experiment
\cite{ThompsonSiggia95}, {\it ii)} the contribution of several competing
secondary structures with comparable free energy but different extension, and,
most importantly, {\it iii)} the fact that several globules are pulled upon in
parallel so that opening one of them may be accompanied by the re-closure of a
neighbouring one, smearing out the jumps in extension that one would observe
from a single globule.

Since we do not aim at giving an accurate prediction of force-extension curves
but rather want to analyze the dynamics in the secondary structure, we will 
circumvent the first two smoothing effects by eliminating thermal
fluctuations: Instead of averaging over all secondary structures with their
appropriate Boltzmann weight (at fixed force) we restrict ourselves to the
structure with the lowest free energy which allows us to get rid of the
entropic fluctuations in the phase space of pairing patterns. Furthermore, we
characterize the extension of a given secondary structure by its fully
stretched end-to-end distance rather than the extension in real space which is
subject to thermal fluctuations.
In this way, we can analyze the direct signatures of sequence disorder as a
succession of sharp jumps in the equilibrium force-extension characteristics.
In an experiment, they would show up as bistabilities
\cite{Rief97,BockelmannThomen02,ChenDill00} that are further smoothed by
entropic fluctuations of the extension. However, we believe that (at low
enough temperatures) thermal effects do not alter the physics in an essential
way apart from blurring the data to a certain extent.

The second order phase transition mentioned above appears in a new light if considered from the point of view of sequence specific response. The critical force separates two qualitatively different regimes: At low force, the chain folds into very few large globular structures that may rearrange dramatically upon an increase of the force under quasi-equilibrium conditions, thus revealing information about folding pathways and the energy landscape. On the other hand, at forces above the threshold the chain organizes into a necklace-like structure with an extensive number of small globules linked by unpaired single strands. The disorder manifests itself only weakly in this phase since the sequence specific response of the individual globules is averaged out when the force pulls on an extensive number of structures in parallel. In the high-force regime the force-extension curves become indeed rather featureless. 

In this paper, we study in detail the critical behaviour in the vicinity of the threshold force. We find numerically that different disordered models belong to the same universality class and derive relations between the critical exponents. Extrapolation of the scaling laws to zero force allows us to obtain the scaling behaviour of various observables in the ground state. In particular, we discuss the scaling of the radius of gyration and the implications for the mean monomer density in real space. Furthermore we characterize the statistical properties of the jump-like events in the low-force and the critical regime, emphasizing the relevance of these results with respect to single molecule experiments. 

The models are introduced in section \ref{model}. In section \ref{homomodel} we review the properties of homogeneous RNA and introduce the relevant structural observables. The algorithm to obtain the force-extension curve for disordered RNA is explained in section \ref{dismodel}, and the results are presented in section \ref{results}. Finally, section \ref{summary} concludes with a summary and an outlook including experimental perspectives. 

\section{Models for RNA}
\label{model}

\subsection{Topological rules for the secondary structure}
We will use a simplified model of RNA as in previous work \cite{Higgs96,BundschuhHwa01b,KrzakalaMezardMuller02}, describing the folding of a sequence of $N$ bases $\{b_i\}_{i=1,\dots,N}$ by its secondary structure, i.e., the list of ordered base pairs $(b_i,b_j)$ with $i<j$, whereby each base can at most be paired to one other base. As usual in the prediction of secondary structure (see e.g. \cite{BustamanteTinoco99}), we do not allow for pseudo-knots, i.e., base pairs $(b_i,b_j)$ and $(b_k,b_l)$ with either $i<k<j<l$ or $k<i<l<j$. In order to account for the importance of the base stacking energy as compared to the covalent pairing energy we do not allow for isolated base pairs that are known to be thermodynamically unstable.

The free energy of a given secondary structure is taken to be the sum of independent local contributions, but instead of using the extensive set of rules established by Turner's group \cite{BurkardTurner99} (as has been done in \cite{GerlandBundschuh01}) we study two simplified models which we believe to reproduce the essential features of the energy landscape and the force-extension curves in particular. 

\subsection{The Models}
The {\it Gaussian model} assigns a random pairing energy $e_{i,j}$ to each base pair $(b_i,b_j)$ where the $e_{i,j}$'s are independent Gaussian variables with zero mean and unit variance. Studies of the glass phase with such a model have shown that it captures well the relevant features of more elaborate models. However, due to the absence of detailed microscopic constraints and the continuous distribution of pairing energies the model approximates fairly well the behaviour of real disordered RNA on a coarse-grained level where the bases $b_i$ correspond to small substrands of the real molecule and the $e_{i,j}$ describe effective pairing affinities between those substrands. As a consequence, the finite size effects are smaller than in a model with detailed sequence dependent interactions.   

The {\it four letters model} starts from a sequence of letters A,C,G,U and assigns a stacking energy to each couple of adjacent base pairs where only Watson-Crick (A-U and G-C) and wobble pairs (G-U) are allowed. We have taken the thermodynamic values used in Zuker's algorithm \cite{Zukermfold} at 37\textdegree. The model differs from the full set of Turner rules mainly by the neglect of entropic costs for hairpin and internal loops and bulges. The inclusion of these terms would re-shuffle the low-lying states without altering the qualitative response to an external force. We therefore expect the results of the four letters model to be representative for real molecules.

Both toy models behave qualitatively in the same way but the finite size effects are stronger in the four letters model [and there is a (trivial) bunching of jump events at higher forces due to the discreteness of the stacking energies which does not occur in the Gaussian model with continuous energies]. For most of the subsequent discussion we will present the results of the Gaussian model. However, we will refer to the more realistic four letters model when we discuss signatures of specific single molecules.

Similar toy models have been considered by many authors to study the low temperature behaviour of RNA \cite{Higgs96,PagnaniParisi00,BundschuhHwa01a,KrzakalaMezardMuller02,MarinariPagnaniRicci02}. There is wide agreement that at sufficiently low temperatures the sequence heterogeneity leads to a glassy phase where the molecule is trapped within a rugged energy landscape whose precise character is still not very well understood, however. 

We note in passing that there could also be a glass transition on the level of the tertiary structure, i.e., the arrangement of a given secondary structure in real space, as has been predicted in \cite{GutinShakhnovich93}. Of course, this aspect of the energy landscape cannot be captured by our description restricted to the secondary structure.

\subsection{RNA under tension}
An external force couples to the system by adding a term $-\vec{f}\cdot (\vec{r}_N-\vec{r_1})$ to the total energy. The end-to-end distance of the molecule $\vec{r}_N-\vec{r_1}$ can be decomposed into a sum of contributions from the globules in the chain and from the single strands linking them. We keep the description simple by adding an energy $-fl$ for each backbone element in a linker strand and for each globular structure within the free part of the chain, see Fig.~\ref{Figmountain}. Thereby, $l$ denotes the (mean) projection of the chain elements onto the direction of the force. This description eliminates by hand the thermal fluctuations due to entropic elasticity. By treating the closing bonds of the globules on the same level as the linker elements we most probably underestimate their contribution to the free length, but we do not expect the physics to depend crucially on these details.

Finally, the total energy of a secondary structure $\mathcal{S}$ reads
\be 
\label{hamilton}
 E_N({\mathcal S})=\sum_{{\rm pairs\, or\, stacks\,}\in {\mathcal S}}e_{{\rm pair/stack}}-fl\cdot n({\mathcal S})
\ee
where $n(\mathcal{S})$ is the extension of the chain in the direction of the force, see Fig.~\ref{Figmountain}.

\section{Review of homogeneous RNA}
\label{homomodel}

We briefly review the results \cite{DeGennes68,Muller02} obtained in the
framework of a homogeneous description of RNA where all pairing or stacking
energies are independent of the interacting bases. This model applies to
homogeneous polymers (periodic self-complementary series $ATAT\dots$ or
$GCGC\dots$) as well as to disordered sequences at high temperatures on a
coarse-grained level. For details we refer the reader to \cite{Muller02}.

\subsection{Radius of gyration}
\label{homomodel:Rg}
\begin{figure}[!]
\centering
\resizebox{0.45\textwidth}{!}{\includegraphics{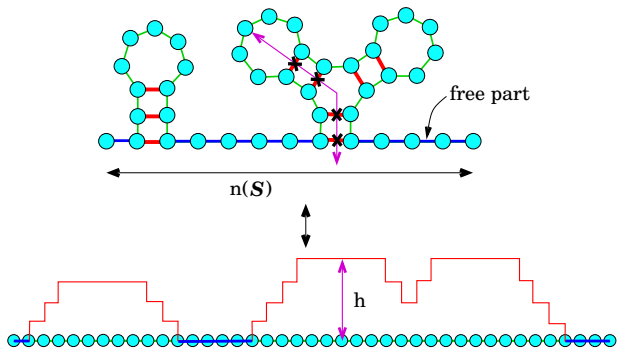}}
\caption{The standard representation (top) of the secondary structure and the corresponding mountain height representation (bottom): Up- or downward steps correspond to bases that are paired down- or upstream in the sequence, respectively. The height of the mountain above a given base equals the number of bonds to be crossed when linking it to the free part of the chain (bases at the bottom of the standard representation) following helices and internal loops, as indicated by the arrow. The average height $\overline{h}$ of the mountain is a measure of the typical distance between peripheral bases. The extension $n({\mathcal S})$ of a secondary structure is taken to be the length of the free part. This equals the number of backbone elements and globules the free part contains.}
\label{Figmountain}
\end{figure}

The radius of gyration $R_g$ of the molecule is an important and experimentally accessible observable. Its leading dependence on the length $N$ of the molecule can be derived most easily from the mountain height representation of the secondary structure \cite{BundschuhHwa01b}, see Fig.~\ref{Figmountain}.

The height of the mountain above the linear sequence indicates the number of bonds that are crossed when one follows the helices and internal loop in the secondary structure from the free part to the considered base. The average height $\overline{h}$ is proportional to the typical distance between peripheral bases within the secondary structure. If the latter is not too dense in real space one may assume that the helices and loops essentially follow a random walk in space, and the radius of gyration is expected to scale as $R_g\sim \overline{h}^{1/2}$. The thermal average of the typical mountain height scales like $N^{1/2}$ since the ensemble of possible mountains is in one-to-one correspondence to a random walk in a half plane, constrained to return to the origin after N steps (possibly with a constant energy gain per up/down step) whose typical excursion is well-known to scale like $N^{1/2}$.

For the homogeneous model we thus deduce $R_g\sim N^{1/4}$ which leads to a monomer density in three-dimensional space scaling like $N/R_g^3\sim N^{1/4}$. As pointed out in \cite{DeGennes68,Muller02} this result can only be valid for small enough $N$ since for large molecules excluded volume effects become essential and cannot be disregarded in the discussion of secondary structure. This makes the usual approaches to RNA folding questionable in the case of large molecules since they are all based on the assumption that tertiary interactions can safely be neglected when the secondary structure is determined. However, we will see below that sequence disorder reduces this problem in that it increases the typical base to base distance in the molecule just sufficiently to avoid too high monomer densities in real space.

Finally, let us introduce the notion of the hierarchy level of a base within the secondary structure which we define as the number of multi-branched loops (loop junction of at least three stems) crossed when following the secondary structure from the free part to the base. It indicates how deep in the tree-like structure of the pairing pattern the base is located. It will be a useful quantity to characterize the size and importance of force-induced rearrangements.

\subsection{Distribution of base-pair lengths}
\label{homomodel:P(l)}

The mountain height is closely connected to the distribution of base pair lengths (defined as $l=j-i+1$ for a base pair $(b_i,b_j)$). In the homogeneous case, the probability distribution $P(l)\sim l^{-3/2}$ (cut off on a scale ${\mathcal O}(N)$) derives directly from the partition function of a molecule with $N$ bases that scales as $Z_N\sim \zeta^{-N}/N^{3/2}$ \cite{DeGennes68,Waterman78}. The distribution of base pair lengths will be modified in the presence of disorder since there will be a tendency to pair bases over longer distances in order to take advantage of favourable pairing energies. This will result in lowering the exponent $3/2$.

The scaling of the average height can be derived from the distribution $P(l)$ by noting that the total area under the mountain can be written as the sum over the length of each base pair,  

\be
	\label{hfromP(l)}
	\overline{h}=\frac{1}{N}\sum_{l=1}^{N} P(l)l\sim \frac{1}{N}\int_1^{N} \frac{l}{l^{3/2}(1-l/N)^{3/2}} dl\sim N^{1/2}.
\ee

\subsection{Critical behaviour under an external force}
\label{homomodel:fc}

Let us now consider the coupling of a homogeneous molecule to an external
force. Up to a critical force $f_c$ the molecule is in a globular state and
its extension remains very small. At high forces the molecule is extended, the
number of bases in the free stretched part being proportional to $N$. The
number of globules is extensive as well, and for forces near to the threshold
it is essentially proportional to the extension since the mean length of the
linker strands in between is only weakly force-dependent. As discussed in
\cite{MontanariMezard01,Muller02} a second order phase transition takes place
at the critical force, and both the extension and the height obey a scaling
law of the form

\begin{equation}
	\label{homscal1}
	h=N^{1/2}\psi_{h}[N^{1/2}(f-f_c)] 
\end{equation}
and 
\begin{equation}
	\label{homscal2}
	\mathcal{L}=N^{1/2}\psi_{\mathcal{L}}[N^{1/2}(f-f_c)]
\end{equation}
as is obtained from a careful analysis of the partition function of a homopolymer under tension, cf., \cite{Muller02}.%
The scaling argument of the functions $\psi_{h}$ and $\psi_{\mathcal{L}}$ indicates the existence of a correlation length that takes the form
\be
\label{correlationlength}
N_c\sim (f-f_c)^{-2}.
\ee
It is related to the size of the largest globules in the chain in the sense that a finite fraction of all bases belongs to globules of size $N_{\rm glob}\ge N_c$. 

The critical force results from a competition between the energy gained from the external force upon increasing the free length of the chain and the corresponding free energy loss. While the latter is of almost purely entropic origin in the homogeneous case (since the extension can increase without reducing the number of paired bases and thus without energy cost) there is also an enthalpic component to be included for disordered models. (In our effective zero temperature study it is even the unique contribution.) Using a Harris-type criterion \cite{Muller02}, one can show that the disorder will only marginally affect the correlation length and the corresponding exponent will remain the same, $\nu^{\rm dis}=\nu^{\rm pure}=2$. At high temperatures, the disordered models turn out to be in the same universality class as homogeneous RNA, while at low temperatures disorder modifies the other critical exponents and renders the globular phase glassy.    

\section{Disordered RNA}
\label{dismodel}

\subsection{Observables}
\label{dismodel:obs}
Since in a homogeneous chain a lot of energetically equivalent states exist that can be smoothly transformed into each other in phase space (by sliding the base pairing pattern) the secondary structure will respond to an applied force in an essentially continuous manner. However, the force-induced unfolding of disordered RNA proceeds in stepwise rearrangements that occur at well-defined forces when thermal smearing is neglected. We will describe these jump-like events statistically, averaging over all events in a narrow force window. The density of jumps, i.e., the frequency of occurrence per unit of force is small below the critical force since the polymer is trapped in metastable states, but it becomes extensive in the high force regime. The relevant characteristics of rearrangements are their sizes as given by the number of bases that change their pairing behaviour and the depth that is best captured by the largest hierarchy level involved in the rearrangement. 


\subsection{Numerical methods}
\label{numerics}

\begin{figure}[h]
\centering
\resizebox{0.45\textwidth}{!}{\includegraphics{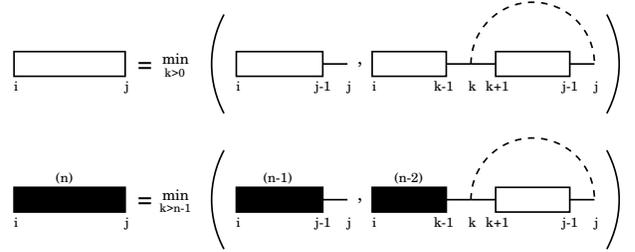}}
\caption{Schematic representation of the recursions over substrand lengths used to calculate the ground state energy $E_0(i,j)$ of the substrand from base $i$ to $j$ (white boxes, top), and the ground state energy $E_L(j;n)$ of the substrand from base $1$ to $j$ constrained to have extension $n$ (black boxes, bottom).}
\label{Figrecursion}
\end{figure}

The force-extension curves of random RNA were obtained numerically by determining all jump events, i.e., the set of forces at which the lowest free energy state changes, together with the corresponding secondary structures. To this end, we first solved recursively the folding problem restricted to the substrand from bases $i$ to $j$ with the standard $\mathcal{O}(N^3)$ algorithm introduced by McCaskill \cite{ZukerStiegler81,McCaskill90}, that yields the ground state energies $E_0(i,j)$ and the corresponding configurations (see Fig.~\ref{Figrecursion}). In a second step, we determined the ground state $E_L(j;n)$ of the substrand from bases $1$ to $j$ constrained to contribute $n$ elements to the free part of the chain (with $n<j$). This can be done using the recursion (see \cite{GerlandBundschuh01} for a similar algorithm)
\begin{eqnarray}
\label{recursion}
E_L(j;n\ge2) = \min &&\Big\{\min_{k=n,\dots,j-2} \Big[ E_L(k-1;n-2)\Big.\Big. \nonumber\\
&&\quad\quad\quad \Big.+e_{k,j}+E_0(k+1,j-1)\Big],
\nonumber\\
&& \quad \Big. E_L(j-1;n-1)\Big\},
\end{eqnarray}
starting from the initial conditions
\begin{eqnarray}
\label{initcond}
E_L(j;0)&=&0,\nonumber\\
E_L(2;1)&=&E_0(1,2),\\
E_L(j>2;1)&=&e_{1,j}+E_0(2,j-1).\nonumber\\
\end{eqnarray}

In order to keep the explanation of the algorithm simple the restriction that isolated base pairs are not allowed has been dropped here and we only treated the case of pairing interactions (as in the Gaussian model). Both stacking energies and the exclusion of isolated base pairs can be taken into account by a minor modification of the algorithm.

The lowest free energy for fixed extension $n$ in the presence of force is given by $E_L(N;n)-f\cdot n$. The extension $n_0$ at vanishing force is the value of $n$ that minimizes $E_L(N;n)$. The forces $f_i$ at which jumps of the ground state configuration occur, together with the respective extensions $n_i$, are iteratively determined from
\be
\label{fis}
f_{i+1}=\min_{n_{i+1}>n_i}\frac{E_L(N;n_{i+1})-E_L(N;n_i)}{n_{i+1}-n_i},
\ee
$n_{i+1}$ being the argument that minimizes the right hand side. (In the case of a degeneracy we chose the largest possible $n_{i+1}$.) 

Once all force intervals and the corresponding lowest free energy states were known, we averaged the single state observables (extension, height, maximal hierarchy level) over all states in a window of width $\Delta f=0.01$, while the characteristics of jump events (density, size, hierarchy level involved) were averaged over all events in a window of width $\Delta f=0.05$. Finally, a quenched average over about $1000$ samples of fixed lengths in the range $N=200-2000$ was performed. 

\section{Results}
\label{results}
\subsection{General properties of globular and extended phase}

\begin{figure}
\resizebox{0.5\textwidth}{!}{
  \includegraphics{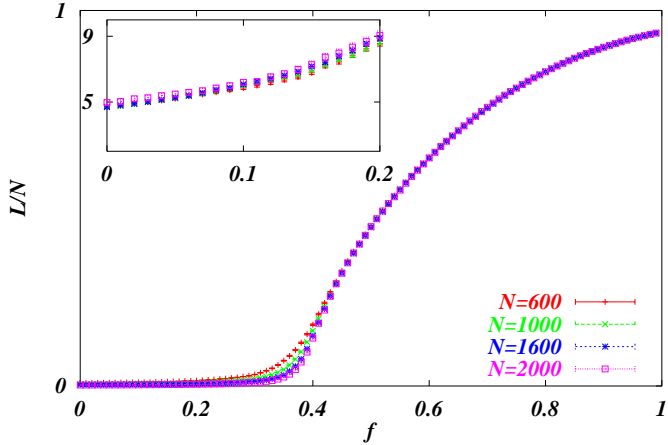}}
\caption{Extension per base as a function of the force (Gaussian model, average over 1000 samples). The extension in the high force regime is clearly seen to be extensive. The inset shows the total extension of the chain in the low force regime. There the extension is very small and the chain typically does not contain more than $3$ globules.}
\label{Fig_phase_l}
\end{figure}

\begin{figure}
\resizebox{0.5\textwidth}{!}{
  \includegraphics{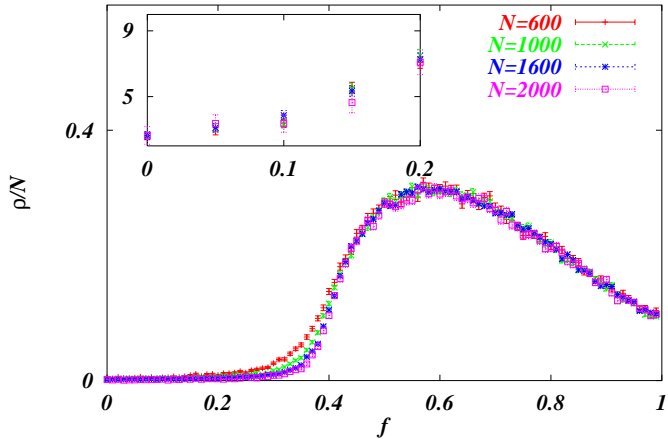}}
\caption{Frequency of occurrence of rearrangements per unit of force, divided by $N$. In the high force regime the rearrangements occur locally and effect mostly single globules. The associated jump forces are uncorrelated and thus the frequency of events is proportional to the number of globules and thus to $N$.  The inset shows the unscaled frequency of occurrence in the low force regime. Up to the critical force $f_c\approx 0.39$ only few jumps occur on average, their number being essentially independent of the size of the molecule.}
\label{Fig_phase_d}
\end{figure}

\begin{figure}
\resizebox{0.5\textwidth}{!}{
\includegraphics{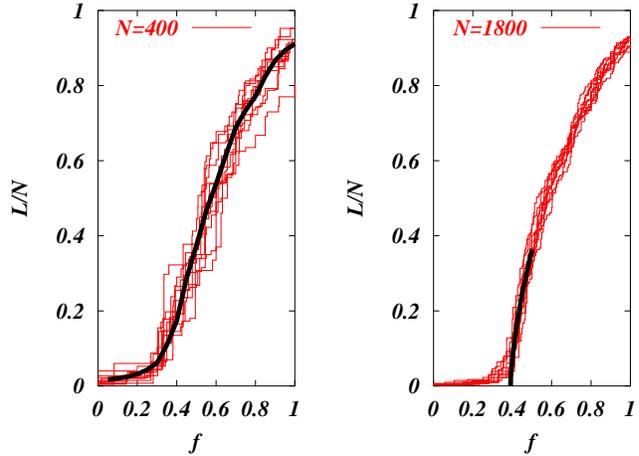}}
\caption{Force-extension plot for $10$ random sequences of the four letters model for $N=400$ (left) and $N=1800$ 
(right). The magnitude of the sample to sample fluctuation decreases 
clearly when
$N$ increases, demonstrating that the system is self-averaging in the extensive phase. The continuous line in the left plot indicates the sequence-averaged extension, while in the right figure we plot the non-linear critical law ${\mathcal L}\sim (f-f_c)^{2(1-\gamma)}$ to be expected in the thermodynamic limit.}
\label{Fig_sse_1}
\end{figure}

In Fig.~\ref{Fig_phase_l} we plot the extension per base, $n(f)/N$, as a function of the force for the Gaussian model. Above the critical force $f_c\approx 0.39$ the chain contains an extensive number of globules, and its free length is proportional to the number of bases $N$. At low forces, the extension is very small and on average there are only very few globules in the chain, independently of $N$. This is also reflected in the frequency of rearrangements, see Fig.~\ref{Fig_phase_d}: Jumps are very rare below the critical force, on average only one or two usually rather important rearrangements take place up to $f=0.2$. The situation changes dramatically above $f_c$ where the force pulls on an extensive number of essentially independent globules in parallel and a typical rearrangement only involves a single globule. In a model with continuously distributed pairing energies, the globules rearrange or break up at well-defined forces that are practically uncorrelated among each other. Thus, the frequency of such events is expected to be proportional to the number of globules and thus scales like $N$. The release of bases, i.e., the increase of extension per jump event is finite and rather small on an experimental scale so that the force-extension characteristics becomes very smooth. This effect is also at the basis of self-averaging of the force-extension curves 
in the extensive phase
as illustrated in Fig.~\ref{Fig_sse_1}. The sample-to-sample fluctuations in the extension per base at a given force decrease with $N$ since the force acts on a large number of globules in parallel and disorder effects are averaged out.

The rearrangements are much more interesting in the low force regime and around the critical force. Far below the critical force a typical rearrangement consists in a large globule breaking up in several (up to  7) smaller substructures. Thereby, the rearrangement is not just superficial, that is, it does not only concern the uppermost levels of the hierarchical tree structure but typically involves a hierarchy level that grows like $N^{0.5}$ as we will see later. It reaches an average of $8$ for $N=1600$ at low forces in the Gaussian model which means that (at least) a cascade of 8 successive helices has to open up (and close differently) during the rearrangement. This involves a complicated pathway in the space of secondary structures, and the equilibration times in this regime of forces will be considerable. One therefore expects to see slow dynamics in real experiments that may reveal interesting information on the intermediate stages of the refolding.

A rearrangement is not necessarily restricted to a single globule but can involve several neighbouring ones. This sort of cooperativity is particularly large slightly below the critical force where on average $2-3$ globules are involved. At higher forces the most frequent events are those where one small globule is opened up and stretched out.

\subsection{Scaling and universality}

Let us now examine the phase transition and the associated critical behaviour in more detail. In analogy to the homogeneous case, Eqs. (\ref{homscal1}) and (\ref{homscal2}), we expect the mountain height and the extension to obey scaling laws 
\be
	\label{hscaling}
	h=N^{\beta}\psi_{h}[N^{1/2}(f-f_c)]
\ee
and 
\be
	\label{lscaling}
	\mathcal{L}=N^{\gamma}\psi_{\mathcal{L}}[N^{1/2}(f-f_c)].
\ee
Note that we used the same correlation length exponent $\nu=2$ as in the homogeneous case as suggested by a Harris-type criterion \cite{Muller02}. The exponents $\beta$ and $\gamma$ may however be modified by the disorder.%
Since forces sufficiently below $f_c$ are irrelevant for the structural properties such as the height $h$, it follows that the scaling function $\psi_h(x)$ tends to a constant for large negative values of $x$. Thus, we can extrapolate (\ref{hscaling}) to $f=0$ and obtain the scaling behaviour in the ground state $h(f=0)\sim N^\beta$. The determination of $\beta$ via the critical scaling allows for a much better control of finite size effects than a simple fit at $f=0$ (as was done in \cite{BundschuhHwa01a}) where small samples are effected by the vicinity of the critical point.

In Eq.~(\ref{hfromP(l)}) we established a general relation between the average height $\overline{h}$ and the probability distribution of base pair lengths, $P(l)$, both in the absence of force. In analogy to the homogeneous case, we expect a power law decay for the probability of long pairings according to $P(l)\sim l^{-\alpha}$, from which we derive the scaling of the average height as $\overline{h}=1/N\sum_{l=1}^{N} P(l)l\sim N^{2-\alpha}$, and hence $\alpha=2-\beta$. Since the disorder favours distant pairings in order to gain energy from particularly well matching substrands, we expect $\alpha$ to be lowered with respect to the homogeneous case (where $\alpha=3/2$).

\begin{figure}
\resizebox{0.5\textwidth}{!}{
  \includegraphics{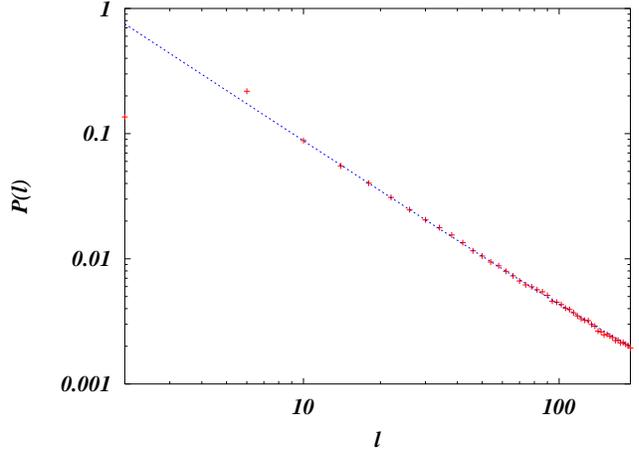}}
\caption{Histogram of base pair lengths at $f=0$, averaged over sequences of length $N=1600$ in the Gaussian model. The curve is the best fit to ${\rm const.} l^{-\alpha}$ which yields $\alpha\approx 1.34$.}
\label{Fig_P(l)}
\end{figure}

\begin{figure}
\resizebox{0.5\textwidth}{!}{
  \includegraphics{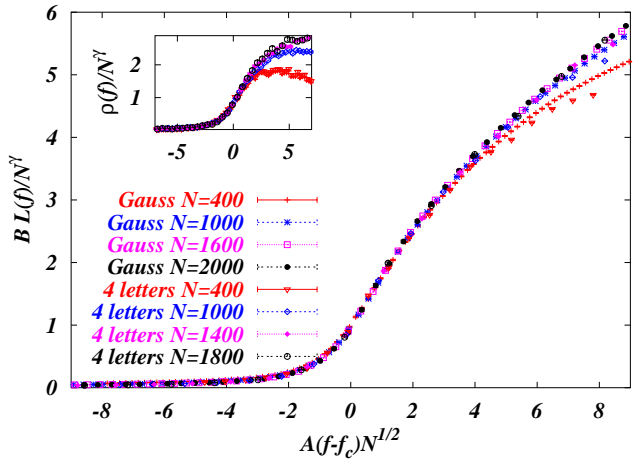}}
\caption{Universal critical scaling of the extension for both the Gaussian and the four letters model. $A$ and $B$ are model-dependent constants. The critical exponent is found to be $\gamma\approx 0.71$. The inset shows a scaling plot of the frequency of rearrangements in the Gaussian model with the same critical exponent $\gamma$.}
\label{Fig_Lscal}
\end{figure}

\begin{figure}
\resizebox{0.5\textwidth}{!}{
  \includegraphics{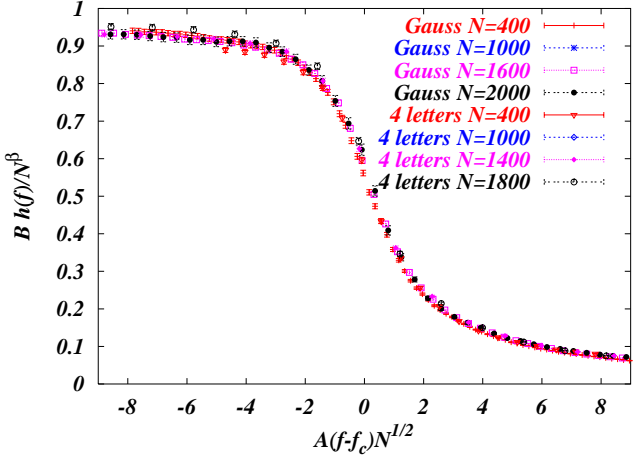}}
\caption{Universal critical scaling of the mountain height $h$ for both disordered models. The constants $A$ and $B$ are the same as in Fig.~\ref{Fig_Lscal}. From the extrapolation to large negative values, the scaling $h(f=0)\sim N^\beta$ can be extracted with $\beta\approx 0.67$.}
\label{Fig_Hscal}
\end{figure}

In Fig.~\ref{Fig_P(l)} we plot the sample averaged probability distribution $P(l)$. The best fit to a power law yields $\alpha\approx 1.340\pm 0.003$. Scaling plots for the height and the extension, with optimized values for $f_c$, $\gamma\approx 0.71\pm 0.02$ and  $\beta\approx 0.67\pm 0.02 $ are shown in Figs.~\ref{Fig_Lscal} and \ref{Fig_Hscal} for both the Gaussian ($f_c=0.395\pm 0.005$) and the four letters model ($f_c=0.39\pm 0.01$). Note that the exponent relation $\beta=2-\alpha$ is very well confirmed by the numerical data. After a global rescaling of the axes with model dependent factors $A$ and $B$ the curves all collapse on a single scaling function. This strongly suggests that the two different disordered models belong to the same universality class, implying that the critical exponents are independent of the specific kind of disorder (but clearly distinct from the values in the homogeneous case). 

In the critical regime, there are numerous globules in the chain, each of which defines a (randomly distributed) force where it will rearrange. The frequency of rearrangements is thus proportional to the number of globules which in turn is proportional to the extension, at least in the critical regime. Therefore the frequency of jumps scales with the same critical exponent $\gamma$ as the extension, as shown in the inset of Fig.~\ref{Fig_Lscal}.


The scaling (\ref{lscaling}) suggests that in the thermodynamic limit the extension grows like $\mathcal{L}\sim (f-f_c)^{2(1-\gamma)}N$ just above the critical force, i.e., exhibiting a rather unexpected non-linear response. Experimentally this effect can only be seen in very large molecules that are sufficiently self-averaging in order not to mask the phase transition by sequence specific signals, see Fig.~\ref{Fig_sse_1}.

\subsection{Base pair lengths and the radius of gyration with disorder}


Following the reasoning in section \ref{homomodel:Rg}  we expect the radius of gyration to scale like $R_g\sim h^{1/2}\sim N^{\beta/2}$. This in turn gives rise to a monomer density in real space scaling like $N/R_g^3\sim N^{1-3\beta/2}\approx N^0$. In contrast to the homogeneous model where this density grows as $N^{1/4}$, sequence disorder leads to more elongated secondary structures that are not increasingly dense. In this sense, disorder re-establishes (at least marginally) the self-consistency of the standard approach to RNA-folding which neglects excluded volume effects (and other tertiary interactions) when optimizing the secondary structure. The latter would not be justified if the individual side chains came into conflict with each other in real space, as would be inevitable for large molecules if $\beta < 2/3$.

The inequality $\beta \ge 2/3$ is only a necessary global condition to avoid too dense structures. However, it is unclear whether it is sufficient to prevent structures with too high {\it local} densities. In any case, since the global condition is just marginally verified in random disordered RNA, we expect that excluded volume effects still play an important, if not decisive, role in large molecules and have to be taken into account in the secondary structure prediction of large molecules.

\subsection{Equality of $\beta$ and $\gamma$}
The critical exponents $\beta$ and $\gamma$ seem to be equal within the error bars. We will argue in favour of their equality by considering statistical properties of the secondary structure at the critical force. 

At $f_c$ the number of globules in the chain is already large ($\sim N^\gamma$) and we assume that the globule sizes $L$ are distributed according to a power law $P(L)\sim L^{-\delta}$ with a cut-off on the scale $N$. This has been confirmed numerically with an exponent $\delta\approx 1.7$. The total number of bases in the chain can be estimated as the sum over all globule sizes, 
\be
\label{allbases}
N \approx N^\gamma\int_1^{ {\mathcal O}(N)}P(L)L\, dL \sim N^{\gamma} N^{2-\delta},
\ee
which implies $\gamma=\delta-1$.

Let us now consider the largest globule in the chain. The scaling with $N$ of the number $L_{\rm max}$ of bases it contains follows easily from extreme value statistics,
\be
	\label{maxsize}
	\int_{L_{\rm max}}^{\infty}P(L) dL={\mathcal O}(N^{-\gamma}),
\ee 
or $N^\gamma/L_{\rm max}^{\delta-1} = {\mathcal O}(1)$. Thus, $L_{\rm max}\sim N$, i.e., the largest globule contains a finite fraction of all bases. We recall that according to Eq.~(\ref{hscaling}) its height scales as $N^\beta$.

\begin{figure}
\resizebox{0.5\textwidth}{!}{
  \includegraphics{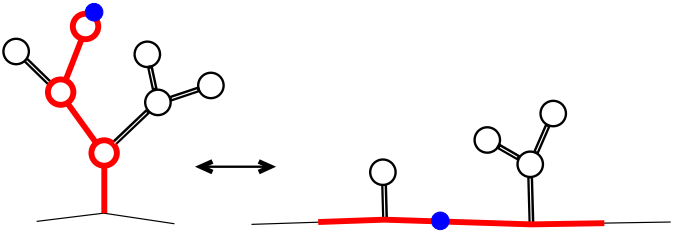}}
\caption{A large structure is split into two halves along the path from the free part to the most distant base (thick line in the left). At the critical force, the cost in pairing energy for this splitting will essentially equal the energy gained from the external force upon stretching (right part). We therefore consider the folding within the full globule to be essentially equivalent to the folding of the same strand exposed to the critical force.}
\label{Fig_fcequiv}
\end{figure}

For the further discussion, we refer to Fig.~\ref{Fig_fcequiv}. The path through the secondary structure from the top most base in the mountain representation down to the free part splits the largest globule into two parts. If we imagine to apply the critical force separately to those to parts (see the right part of Fig.~\ref{Fig_fcequiv}) we would expect their new folding to be qualitatively the same as in the intact globule.
This is because the bases marked by the thick lines in Fig.~\ref{Fig_fcequiv} will gain an equivalent energy from exposure to the critical force as they gained from the pairing to the bases of the other strand. This simply reflects the equality of the free energy per base in a globular structure and in the free part at the critical force. The height of the original globule ($\sim N^\beta$) must therefore scale in the same way as the extension of one of the two halves subject to the critical force. For the latter Eq.~(\ref{lscaling}) implies $h\sim L_{\rm max}^\gamma\sim N^\gamma$, and hence the equality $\beta=\gamma$.

This picture is slightly too simplistic, for one expects that a certain
fraction of the side structures of the globule have a smaller (local) critical
force and would open up when the substrands are exposed directly to the
(global) critical force. However, the unzipping of such weakly bound side
structures will by far not be complete. In a first step, the closing helix will
be unzipped up to the first internal loop, and the subsequent structures
attached to that loop will become directly exposed to the critical force. Again, a
fraction of them will be unstable and rearrange. If one assumes this fraction
to be constant (and not too large) on all hierarchy levels, the number of
opening base pairs in a side structure is finite. We thus conclude that the
total number of opening base pairs in the complete substrand is at most
proportional to the number of side structures, and thus scales like the height
$h\sim N^\beta$ of the globule. The new extension ${\mathcal L}$ of the
substrands therefore still scales like $\sim N^\beta$, though with a larger
prefactor.

We expect finite size effects to be responsible for the slight difference in the numerical values found for $\beta$ and $\gamma$. The partial opening of side structures in the above argument, is naturally cut off at the maximal hierarchy level in the considered secondary structure. The latter is still rather small for the molecule sizes we studied numerically. The ratio ${\mathcal L}/h$ between the extension of the substrands and the height of the original globule becomes slightly larger when more hierarchy levels are available, but it will saturate at sufficiently large sizes. This might be at the basis of the numerical trend to find $\gamma>\beta$.%

\subsection{Characteristics of force driven jumps}
\label{jumps}
Let us now analyze in more detail the single rearrangements and their statistics. The depth of a rearrangement is best characterized by the maximal hierarchy level involved. We note that the total number of hierarchy levels in a secondary structure scales in the same way as the height ($\sim N^\beta$) since a single level always contains a bounded number of bases.

\begin{figure}
\resizebox{0.5\textwidth}{!}{
  \includegraphics{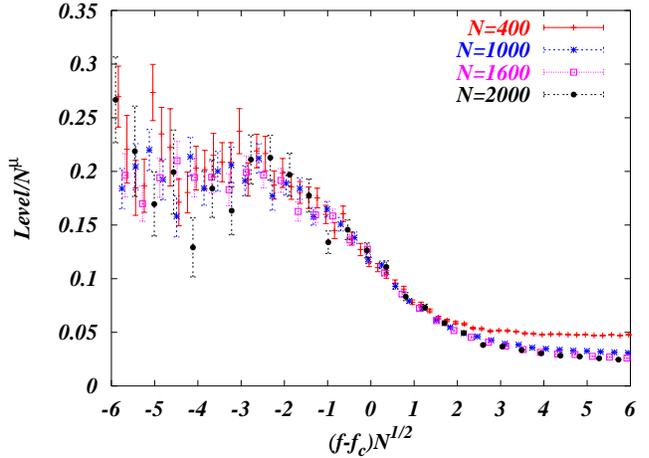}}
\caption{Scaling plot of the hierarchy level involved in a rearrangement, averaged over events in a force window $\Delta f=0.01$. The scaling at $f=0$, $N^{\mu}$, with $\mu=0.5$, is smaller than that of the maximal hierarchy level in the structure ($N^{0.65}$). However, it shows that a substantial fraction of the globule is involved in a rearrangement.}
\label{Fig_phase_hierlevel}
\end{figure}

To find the scaling of the average level involved in a jump we again resort to a critical point analysis, see Fig.~\ref{Fig_phase_hierlevel}, and extrapolate the scaling function to $f=0$. The scaling is less neat than for other observables since finite size effects are rather important. Furthermore the statistics of jump events is sparse in the low force regime which causes the large error bars. Nevertheless an approximate increase of the average level as $N^{0.5}$ can be established, showing that usually a considerable fraction of a globule takes part in rearrangements. The latter will be far from trivial, giving rise to long equilibration times and ageing effects in the low force regime, even in molecules of moderate size.
 
Another important characteristic of a rearrangement is the fraction of bases that change their pairing behaviour. This includes the bases that were paired before the rearrangement and are unpaired afterwards, or vice versa. This quantity measures a kind of phase space distance between the secondary structure before and after the jump. We denote it by $1-q$ where $q$ is the overlap between the secondary structures which bears some resemblance to the overlap defined in spin glasses \cite{Higgs96}. 

\begin{figure}
\resizebox{0.5\textwidth}{!}{
\includegraphics{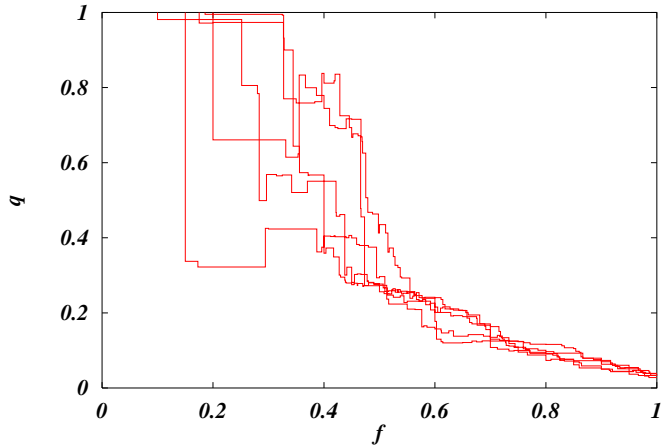}}
\caption{Overlap between the secondary structure at force $f$ and the ground state ($f=0$) for five sequences with $1800$ bases in the four letters model. The amplitudes of jumps are very important in the low force phase, where the behaviour is strongly sequence-dependent. In the high force regime the overlap with the ground state is not particularly meaningful. Very stable structures that have resisted any rearrangement up to that point are finally broken up by the force. This happens in an almost continuous manner since the individual structures are their respective rupture forces are almost randomly distributed.}
\label{Fig_sse_2}
\end{figure}

Macroscopic rearrangements correspond to phase space distances $1-q\sim {\mathcal O}(1)$. At low forces, the relative probability of occurrence for such events is found to decrease with the system size as $\langle 1-q \rangle\sim N^{-\sigma}$, where $\langle \rangle$ denotes the average over rearrangement events in a small force window and the subsequent average over disorder realizations. The exponent $\sigma\approx 0.25\pm 0.05$ is relatively small, and thus, in RNA of moderate sizes macroscopic rearrangements are rather frequent. We illustrate this point in Fig.~\ref{Fig_sse_2} where we plot the overlap of the secondary structure
at a given $f$ with the groundstate ($f=0$) structure for five randomly chosen sequences in the four letters model. At low force, large changes on the scale of the system size are quite common, but of course, the behaviour differs a lot from sample to sample. The high force regime is far less interesting, being dominated by the rupture of small globules.

In the critical regime the changes in the secondary structure are also quite dramatic since a lot of rather important rearrangements occur in a small force window. Indeed, the phase space distance $1-q(f;f+df)$ between the equilibrium secondary structures at the forces $f$ and $f+df$ (for $df$ fixed) increases approximately as $N^{0.3}$ provided both $f$ and $f+df$ belong to the critical regime (for fixed $df$, this restricts $N$ to values such that $N df^2$ is still small). This suggests that hysteresis effects in cycling force experiments are particularly strong in the vicinity of the critical point.

Force-extension experiments clearly probe the energy landscape of RNA. However, it is difficult to relate the statistics of the jump events, and in particular the overlap between the secondary structures before and after the jump, to global properties of the energy landscape. If the latter is characterized by a unique size-dependent energy scale $E(N) \sim N^\theta$, as proposed in \cite{KrzakalaMezardMuller02,MarinariPagnaniRicci02}, one might conjecture that at low forces the probability distribution of large rearrangements in a chain of length $N$ follows a scaling law $P_N(q)= E(N)\tilde{P}((1-q)E(N))$. However, the data is not consistent with such a simple ansatz. This is probably due to the fact that the force does not couple to the secondary structure via a bulk perturbation (as in the $\epsilon$-coupling method used in \cite{KrzakalaMezardMuller02,MarinariPagnaniRicci02}), but only via the extension. This seems to slightly favour successions of smaller rearrangements instead of single large jumps.

Let us recall at this point that we completely neglected thermal fluctuations,
and thus all jump events are infinitely sharp and occur at well defined
forces. In reality such transitions are smoothed out over a small
force window, and if the density of rearrangements is high, the individual
peaks cannot be distinguished anymore. This is true in particular for the
extensive phase, and lies at the basis of the observed smoothness of
experimental force-extension curves. In that regime it proves to be better to
identify transitions between competing foldings via increased fluctuations
\cite{ChenDill00,GerlandBundschuh01}. However, at low forces, thermal effects are far 
less important since the rearrangements are rare, and we expect the present analysis to give a good qualitative picture of typical experimental curves.
Of course, itdoes not reproduce correctly the detailed characteristics of force-extension curves, but it
provide a rather quantitative insight into the nature of the energy landscape as described
by the size and frequency of occurrence of rearrangements, in particular in the low force regime. 

\section{Summary and Discussion}
\label{summary}
We have studied the force-induced unfolding of random disordered RNA under equilibrium conditions, whereby we have eliminated thermal fluctuations by restricting ourselves to the secondary structure with the lowest free energy. The disorder effects on the second order opening transition have been analysed in detail and the phases at low and high force have been characterized in terms of their different structural properties and response to the force. The critical behaviour of the opening transition was found to be modified with respect to the homogeneous case and strong evidence for the universality of different disorder models has been provided. The extrapolation of scaling laws to vanishing force has allowed to determine the scaling of several observables with $N$ in a clean manner.

The statistics of rearrangement events at low forces have been characterized. While in the thermodynamic limit global macroscopic 
rearrangements of the secondary structure are ruled out since their probability decreases as $N^{-0.25}$, macroscopic jump events remain possible up to quite large system sizes.

From the experimental point of view the low force regime is certainly the most interesting since the transformations in the secondary structure required to reach equilibrium are rather complex and will proceed in several steps, possibly passing through metastable intermediates. It will be interesting to observe the approach to equilibrium after slightly changing the force to a new fixed value. Alternatively, cycling force experiments should exhibit interesting hysteresis effects as have been observed in the closely related DNA un- and rezipping \cite{BockelmannThomen02}. The critical regime is particularly suited for this kind of experiments since the overlap between equilibrium secondary structures decreases very rapidly as the force is varied.
 
Such experiments might also shed some light on the energetic barriers $B(N)$ between metastable states in globules of size $N$. Those are related to typical equilibration times $\tau(N)$ via $\tau(N) \sim \exp[B(N)/kT]$. On the theoretical side, those barriers are not very well understood yet. In the spirit of the scaling picture \cite{KrzakalaMezardMuller02,MarinariPagnaniRicci02}, one would expect $B(N)\sim E(N)\sim N^{\theta}$, with $\theta\approx 0.15-0.35$, whereas the analysis of barriers in the ground state ensemble of a degenerate toy model \cite{Higgs98} 
as well as dynamic simulations \cite{Fernandez90,FernandezShakhnovich90} of RNA-folding 
rather point towards barriers growing like $B(N)\sim N^{1/2}$. Further work is needed to clarify this important issue.

\section{Acknowledgments}
M.~M{\"u}ller and F.~Krzakala acknowledge a fellowship from the MRT.
The LPTMS is an Unit\'e de Recherche de l'Uni\-versit\'e Paris~XI 
associ\'ee au CNRS.

\bibliographystyle{prsty}
\bibliography{../Bib/references}

\end{document}